# Prediction of tone detection thresholds in interaurally delayed noise based on interaural phase difference fluctuations

Mathias Dietz[a)], Jörg Encke, Kristin I Bracklo, Stephan D Ewert

Department für Medizinische Physik und Akustik and Cluster of Excellence "Hearing4all", Universität Oldenburg, 26111 Oldenburg, Germany, Universität Oldenburg, 26111 Oldenburg, Germany

[a)] Author to whom correspondence should be addressed.
Email: m.dietz@uol.de

# Abstract

Differences between the interaural phase of a noise and a target tone improve detection thresholds. The maximum masking release is obtained for detecting an antiphasic tone ($S_\pi$) in diotic noise ($N_0$). It has been shown in several studies that this benefit gradually declines as an interaural delay is applied to the $N_0S_\pi$ complex. This decline has been attributed to the reduced interaural coherence of the noise. Here, we report detection thresholds for a 500 Hz tone in masking noise with up to 8 ms interaural delay and bandwidths from 25 to 1000 Hz. When reducing the noise bandwidth from 100 to 50 and 25 Hz, the masking release at 8 ms delay increases, as expected for increasing temporal coherence with decreasing bandwidth. For bandwidths of 100 to 1000 Hz, no significant difference was observed and detection thresholds with these noises have a delay dependence that is fully described by the temporal coherence imposed by the typical monaurally determined auditory filter bandwidth. A minimalistic binaural model is suggested based on interaural phase difference fluctuations without the assumption of delay lines.

# I. INTRODUCTION

The human binaural system can exploit differences between the interaural phase of a masker noise and a target tone to improve detection thresholds [1]. The maximum masking release is obtained for detecting an antiphasic tone ($S_\pi$) in diotic noise ($N_0$). It has been shown in several studies that this benefit gradually declines as an interaural time difference (ITD) is applied [2, 3]. Two different hypotheses have been proposed to account for the decline.

One hypothesis, proposed by Langford and Jeffress [2], attributed the reduction of the binaural masking level difference (BMLD) with increasing noise delay to what they referred to as "interaural correlation of the noise". A more contemporary wording for their quantity is "normalized cross-correlation coefficient", i.e. the value of the normalized cross-correlation function at $\tau = 0$.

The other hypothesis, also proposed by Jeffress [4], but before the above, is that using internal time delays ("delay lines") the auditory system has access to more of the cross-correlation function than just the cross-correlation coefficient at $\tau = 0$. Such circuitry has indeed been found in the barn owl, where left and right sided inputs propagate along counterdirected axons [5]. Coincidence detecting neurons along the axonal delays effectively cross correlate the inputs at different values of $\tau$. An ideal delay line could perfectly compensate any external noise delay ($\tau$=ITD) by an opposed internal delay ($-\tau$), allowing for maximum BMLDs even at large noise delays [6]. It is, however, reasonable to assume that such a compensation mechanism introduces errors for increasing internal delays. This increase in error is commonly simulated as a decrease of the density of correlating elements with increasing internal delay. This relationship is captured by the p($\tau$) function [7, 8]. Based on this second hypothesis, models use p($\tau$) as a fitting parameter, i.e. they estimate the delay line length and potency from the decline of the BMLD with noise delay [7-10].

For the first "cross-correlation coefficient"-based concept, however, this degree of freedom does not exist. In this case, the cross correlation, or more generally speaking the complex-valued temporal coherence of the analytical signal $\gamma$ [11], $\gamma(\tau) = \frac{\langle n(t+\tau)n^*(t)\rangle}{\langle |n(t)|^2\rangle}$ is solely determined by the spectrum of the noise [2, 6] and is proportional to the inverse Fourier transform $F^{-1}$ of its power spectral density *n'* (Wiener–Khinchin theorem):

$$\gamma(\tau) \propto F^{-1}(n'(f)). \qquad (1)$$

This means that the bandwidth of the input signal determines the decay of $|\gamma(\tau)|$: the broader the spectrum, the shorter the temporal coherence. The maximum bandwidth observed by the binaural system is limited by some form of band-pass filter. As a consequence, the filter bandwidth effective at the input to the binaural interaction ultimately dictates how binaural unmasking depends on the noise delay. As a first hypotheses, it seems reasonable to assume that the effective bandwidth at the binaural stage matches those from monaural estimates (e.g. 79Hz ERB at 500Hz [12]). Some previous studies indeed already tried to determine the effective bandwidth based on binaural unmasking data. Rabiner et al. [6] found that their experimental data could be best accounted for using an 85-Hz wide (at -3 dB) triangular filter. Langford and Jeffress [2] coarsely estimated a 100 Hz bandwidth but without specifying the filter shape and bandwidth definition. Both estimates are a little larger but close to the monaural estimates.

The goal of this study is, to revisit if and to what extent the decline in binaural unmasking as a function of noise delay can be explained solely based on the decline of temporal coherence in a simple, "minimalistic" binaural model. The model uses fluctuation of the interaural phase difference [13, 14] resembling a physiologically plausible feature that might be extracted from a neural representation of the signal [15]. This physiologically inspired IPD metric is directly related to the correlation coefficient i.e. the degree of coherence [16, 17]. Early attempts [2, 6] connecting noise delay and temporal coherence appear promising, but have not been followed up, to test if filter bandwidths and filter shapes that are most commonly used in more recent models can quantitatively account for the data. If the simulated decline is faster than the experimental decay, delay lines have to be in operation, compensating for the external delay and thus increasing the coherence at the level of binaural interaction.

To complicate the argument, the effective processing bandwidth of the binaural system has previously been discussed controversially, with some studies suggesting it is larger than the monaural filter bandwidth, at least for certain complex maskers [8, 18, 19]. However, there is growing consensus that at least for the simplistic band-pass filtered noise investigated in the present study the binaural filter bandwidth is not wider [9, 10, 20-23].

## II. Experiment

### A. Participants

Ten young normal-hearing volunteers (21-33 years, median 24 years, 5 male, 5 female) were recruited – all university students. Most subjects had some experience in speech in noise tests. To our knowledge, no subject had prior experience in dichotic tone-in-noise detection. All subjects received at least 90 min of training prior to data collection. All audiometric thresholds were equal to or less than 15 dB HL from 125 to 10,000 Hz, pure tone averages less or equal 5 dB HL, and differences across ears did not exceed 5 dB in both pure tone average and at 500 Hz. One reason for these relatively strict inclusion criteria is that it has been shown recently that subjects with a slight (sub-clinical) hearing loss have a reduced binaural release from masking [25]. At the later test frequency of 500 Hz, no subject had a threshold > 5 dB HL in either ear. Subjects were compensated for their time.

### B. Apparatus

Stimuli were generated digitally at a sampling rate of 48 kHz in MATLAB (MathWorks, Natick, MA, United States) using the AFC software package [26] for MATLAB and presented via a Fireface UC USB sound card, over a Sennheiser (Wedemark, Germany) HDA 650 circumaural headphone, calibrated at 500 Hz. The subjects were seated in a double-walled sound booth.

### C. Stimuli

Gaussian noise was band-pass filtered by cutting out spectral components outside the pass-band. All noises were arithmetically centered at 500 Hz and the bandwidths were 25, 50, 100, 150, 200 or 1000 Hz. Noises had a duration of 380 ms including 20-ms raised cosine on- and offset ramps.

The noise level was kept at a constant spectrum level of 45.5 dB relative to 20 µPa. Fully correlated noises were presented with interaural delays ($\tau$) of 0, 2, 4, or 8 ms, or interaurally uncorrelated. The delay was applied prior to gating. Delays were chosen in multiples of the cycle duration at the 500 Hz center frequency, ensuring zero interaural phase difference (IPD) at 500 Hz.

Target tones had a frequency of 500 Hz and a duration of 300 ms, again including 20-ms raised cosine on- and offset ramps. Tones were always presented temporally centered in the noise. They were either interaurally in phase ($S_0$) or antiphasic ($S_\pi$).

### D. Procedure

A 3-interval, 3-alternative forced choice procedure was employed, with two noise-only reference intervals ($N_\tau$) and one target interval including both signal and noise. Subjects selected an interval by pressing the respective number key on a computer keyboard. Feedback was provided.

The signal level, initially 65 dB SPL, was adaptively changed in a 2-down, 1-up staircase procedure, aiming at the 70.7% correct rate [27]. The step size of 4 dB was reduced to 2 and 1 dB after the second and fourth reversal, respectively. After a total of 10 reversals each run was terminated and the average was taken across the last 6 reversals.

$S_0$ and $S_\pi$ conditions were separated into two independent experiments presented in blocks but without a specified order. For $S_0$ conditions only noise delays of 0 and 8 ms as well as uncorrelated noise were measured. For both the $S_0$ and $S_\pi$ conditions the same measurement order principles were applied: For each randomly chosen bandwidth block all noise delays were measured once in random order. To allow for an "acclimatization" to the new bandwidth, two training runs were included at the beginning of each bandwidth block. Once all conditions, i.e. all six bandwidth blocks were measured, the procedure was repeated with new random orders until all conditions were measured 4 times. The study was approved by the Ethics committee of the University of Oldenburg.

## E. Results and Discussion

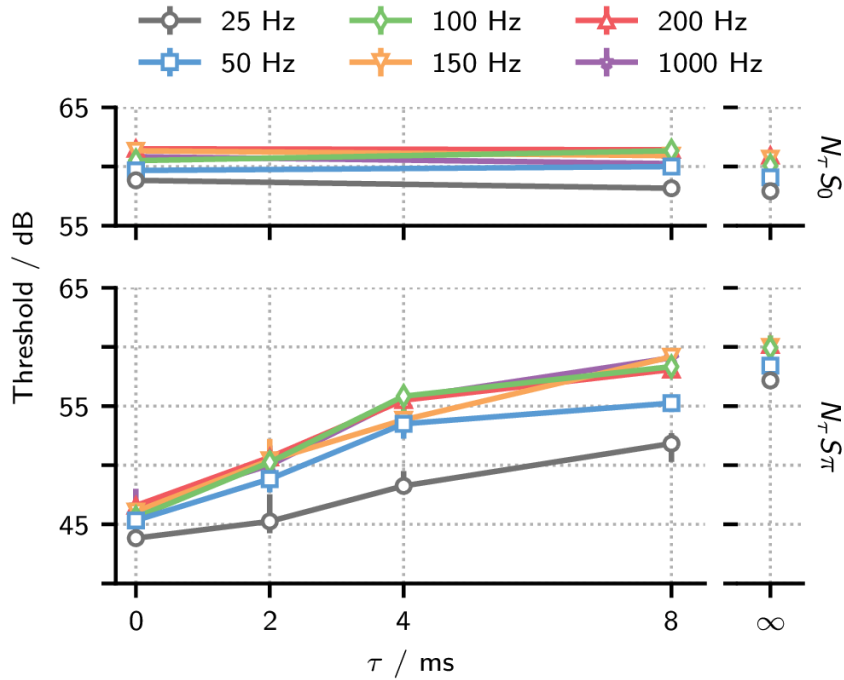

**Fig. 1:** $N_\tau S_0$ detection thresholds (upper panel) and dichotic $N_\tau S_\pi$ thresholds (lower panel) as a function of noise delay in ms. The separated data points at the right-hand side are for uncorrelated noise ($N_u$). Different symbols (color online) are for the different noise bandwidths ranging from 25 Hz to 1000 Hz.

Three of the ten listeners were not able to obtain $N_0S_\pi$ thresholds below +8 dB above the masker spectrum level in the 100-Hz bandwidth noise, while the seven other subjects had thresholds less or equal +2 dB in their formal measurements. Comparable thresholds from other studies are in the range of -3 dB [20] to + 3 dB [3]. Two of the less-sensitive listeners were further tested in a sensitivity-optimized threshold ITD task. Their >100 µs thresholds were larger than any of 52 untrained listeners [28] and thus considered outliers. 3 of 10 listeners with excellent audiograms performed quite poor and in fact worse than the group average of less-sensitive listeners in [25]. To be able to compare the data to previous studies and to meaningfully apply statistics based on normally distributed values, only data from the other seven subjects are reported.

The experimental data is shown in Fig. 1 as a function of the interaural delay from 0 to 8 ms and for interaurally uncorrelated ($N_u$) noise. Symbols depict the median detection thresholds across the seven subjects and the error bars the interquartile range. Different symbols are used for different noise bandwidths. The lower panel shows thresholds for the $S_\pi$ condition. $N_0S_\pi$ thresholds (left-hand data points) were similar to the masker spectrum level and for large bandwidths virtually identical to a large and consistent body of literature [20, 29]. As expected, $S_\pi$ detection thresholds increase with increasing noise delay. The thresholds obtained with 100 to 1000 Hz wide noise are virtually identical with each other and to those obtained by Langford and Jeffress [2]. Smaller bandwidths of 50 and 25 Hz result in slightly lower $N_0S_\pi$ and $N_uS_\pi$ thresholds and increased slower with increasing noise delay. Qualitatively, all these observations were expected as a direct consequence of the increasing temporal coherence in the acoustic stimulus for decreasing bandwidth. For narrow bandwidths up to about 100 Hz, previously reported $N_0S_\pi$ are less consistent across studies. The $N_0S_\pi$ thresholds of the present study are lower than those reported by Bernstein and Trahiotis [3], similar to those by van der Heijden and Trahiotis [8], but overall higher than in the majority of studies focused on thresholds of highly trained listeners obtained with diotic masker [20, 30]. We speculate that mixing in experimental conditions with interaural delay makes it harder for the subjects to fully train on the particularly subtle cues with narrow-band, i.e. tonal maskers and zero delay.

$S_0$ thresholds are shown in the upper panel of Fig. 1. These data were not in the focus of the current study and was recorded only for delays of 0 and 8 ms as well as for the uncorrelated noise to be able to estimate the observed BMLD. The BMLD is on average 14.8 dB without noise delay independent of bandwidth and is reduced to about 6.3 dB and 4.7 dB for 25-Hz and 50-Hz bandwidth at 8-ms noise delay, respectively. For larger bandwidths the BMLD at 8-ms noise delay is only about 2 dB, in line with [2].

To further assess the effect of bandwidth in the $N_\tau S_\pi$ data, a two-way repeated-measures ANOVA [noise delay (5) x bandwidth (6)] was performed, showing a significant main effect of noise delay [$F(4, 24)=464.30$, $p<0.001$], bandwidth [$F(5, 30)=21.53$, $p<0.001$], and a significant interaction [$F(20, 120)=6.47$, $p<0.001$]. Post-hoc pair-wise comparisons (Bonferroni corrected) for the marginal means showed that all noise delays were significantly different ($p<0.01$) from each other. For the bandwidths, only the data for 25 Hz was significantly different from all other bandwidth ($p<0.05$), and the data for 50 Hz was significantly from 200 Hz ($p<0.05$). At the largest delay of 8 ms, 25 Hz was different ($p<0.05$) from all other bandwidths, except for 50 Hz, and 50

Hz was different (p<0.05) from all other bandwidths except for 25 and 200 Hz. Such deviations in the pattern of differences from the marginal mean and between the different delays are the source of the significant interaction term. Taken together, the 25-Hz data, and the 50-Hz data to a lesser degree, are different from the data for the other bandwidths, while the data for 100 Hz bandwidth and above show no significant differences.

## III. Model predictions

### A. Model description

The front end of the model employed here is essentially identical to the IPD model [13, 14] and illustrated in Fig. 2a. Peripheral band-bass filtering is mimicked with a 4$^{th}$-order Gammatone filter centered at the target frequency of 500 Hz, with an equivalent rectangular bandwidths of 79 Hz [12] in the implementation of [31]. Haircell processing is coarsely modelled by half-wave rectification, compression by taking the signal to a power of 0.4 and subsequent low-pass filtering with a 5$^{th}$ order Butterworth filter with a 770 Hz cutoff frequency). For IPD extraction, the phases of left and right signal must be known by definition. The phase is extracted from the (unipolar) haircell representation by applying a second, broader Gammatone filter (2$^{nd}$ order, 167 Hz bandwidth), referred to as temporal fine-structure (TFS) filter, again centered at 500 Hz. From the complex-valued output of this TFS filter, g(t), the argument is the phase. The TFS filter effectively reverts some effects of the haircell nonlinearity, including turning the unipolar signal into a bipolar signal again. In principle, for the purpose of the present study, the phase could have been obtained directly from the first Gammatone filter, however, the haircell stage and the TFS filter were kept as in the IPD model [13, 14] to stay in the conceptual framework of auditory pathway models.

The instantaneous IPD, $\Delta\varphi(t)$, can now be derived by subtracting the phases from the left and the right signal, or, equivalently, by first multiplying the left signal and the complex conjugate of the right signal and then taking the argument from the product. A phase jitter, $X_{\Delta\varphi}$, in the form of Gaussian noise is added to the IPD as a limiting factor of binaural sensitivity, qualitatively corresponding to the time equalization jitter introduced by Durlach [32] or a combined monaural and binaural time jitter [32].

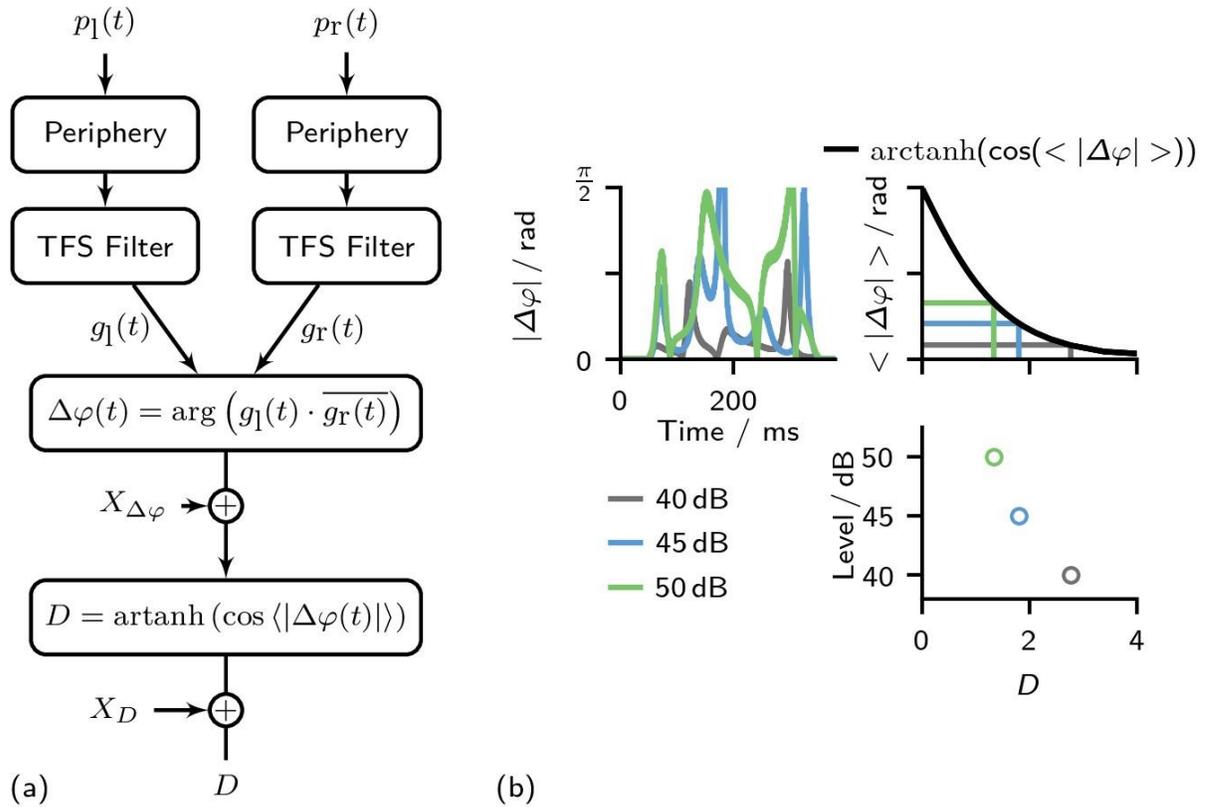

**Fig. 2:** (a) Schematic of the proposed model. (b) Example of the decision stage for $N_0S_\pi$ stimuli at different signal levels. The top left shows the absolute value of the instantaneous IPD over time. Increasing the target level (with an IPD of π) increases fluctuations and thus the mean absolute IPD (y-axis top right). Taking the cosine of the mean IPD results in smaller values with increasing fluctuation. The decision variable D results directly from the cosine, subject to a Fisher's-Z transform, so that lower values indicate a stronger signal prevalence (bottom right).

Adding an $S_\pi$ tone to more intense diotic noise causes the instantaneous IPD to fluctuate around zero. This fluctuation has previously been suggested as the detection cue [13, 17, 34, 35]. In more general terms, it is assumed in the model that the target can be detected if the average fluctuation of the IPD in the target interval can be discriminated from the average fluctuation of the IPD in the noise-alone intervals.

In the current study, the long term average IPD of both target and reference intervals are always zero: $\langle \Delta\varphi \rangle = 0$. It was therefore decided to simplify the model to only detect deviations from zero IPD, i.e. temporal averaging of the modulus across the entire observation interval: $\langle|\Delta\varphi|\rangle$. This value ranges from 0 in case of a diotic stimulus and no internal noise, to $\pi$ for an interaurally antiphasic stimulus. For interaurally uncorrelated noise an average value of $\pi/2$ is obtained, resulting from a uniform distribution of $|\Delta\varphi|$ in the range 0 to $\pi$ in that case. A cosine mapping, projects the values to the interval 1 for no IPD to -1 for an antiphasic stimulus, and to 0 for an interaurally uncorrelated noise. In this simplified version, which can only compare intervals with no offset IPD, the internal variable $\cos\langle|\Delta\varphi|\rangle$ is practically identical to the interaural correlation coefficient, commonly used for similar purposes [10, 29]. This term is then Fisher-Z transformed (Fig. 2b), again identical to the comprehensive binaural detection model by Bernstein and Trahiotis [29]. Last, a detector noise $x_D$, is added, yielding the decision variable D.

$$D = arctanh(\cos\langle|\Delta\varphi|\rangle) + x_D \qquad (4)$$

The *arctanh* Fisher-Z transformation expands differences near 1 and -1 and has previously been employed for correlation-based decision metrics [36] as well as for dichotic tone in noise detection [28]. In combination with the fixed-variance detector noise, an increased sensitivity to interaural correlation differences near 1 and -1 is obtained in combination with decreased sensitivity close to 0 (uncorrelated) as observed in, e.g. [37].

The model back-end is an artificial observer that retrieves the same three intervals in the same adaptive procedure as the listeners and selects the interval with the smallest decision variable D [9, 38]. The artificial observer ran 100 runs in each condition.

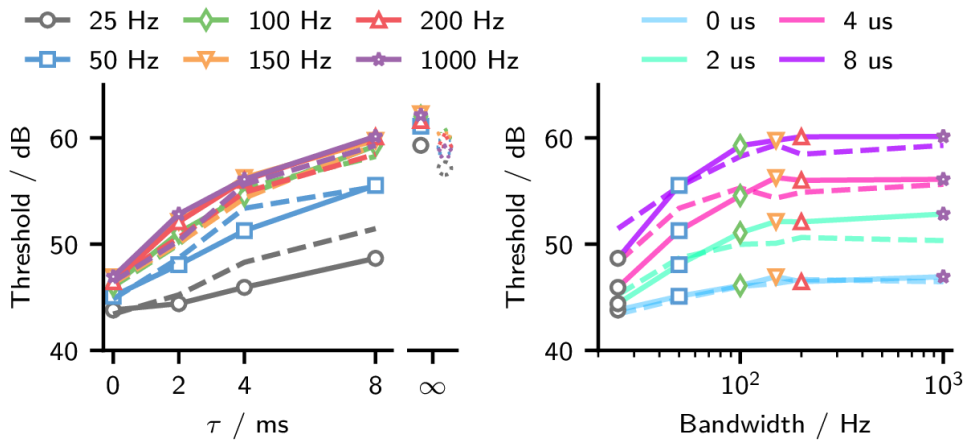

Fig. 3: Model predictions (solid lines with symbols) and experimental data (dashed lines). In the left panel, thresholds are plotted as in Fig. 1 and in the right panel, the same data is plotted

as a function of noise bandwidth with different lines representing the different noise delays. No error bars are shown for the simulations, because the standard error of the mean was < 0.6 dB for all conditions.

## B. Model simulations

Figure 3a shows the model results for the different noise bandwidths as a function of noise delay and for uncorrelated noise, whereas Fig. 3b shows the same data but as a function of noise bandwidth. For comparison, the experimental data is plotted in the background as dashed lines. In the model, the thresholds obtained at 100 Hz bandwidth (diamonds) still differ slightly from the 1000 Hz condition. Both absolute values and the frequency-dependent increase with noise delay are quite accurately captured by the model. Only at 50 Hz bandwidth for 4 ms delay and at 25 Hz the model underestimates the thresholds by up to 3 dB. Diotic conditions were not modeled, given that monaural cues are not captured by the purely binaural model. This also explains that the model thresholds are slightly too high for uncorrelated noise (right-hand side of Fig. 3a): subjects apparently use a combination of weak binaural cues and weak monaural cues in these conditions.

The two internal noise parameters influence the model output in the following way: The internal IPD noise ($X_{\Delta\varphi}$) mostly determines the threshold at τ=0 and with this the slope and curvature of the threshold functions for increasing delay. The decision noise ($X_D$) determines the overall performance, effectively parallel shifting the functions towards higher or lower thresholds. The standard deviations ($\sigma_{\Delta\varphi}, \sigma_D$) of the two noises were determined by first adjusting $\sigma_D$ to fit the data of the $\tau = 8$ ms condition and consecutively adjusting $\sigma_{\Delta\varphi}$ to fit the data at $\tau = 0$ ms which resulted in $\sigma_{\Delta\varphi} = 0.3\, rad, \sigma_D = 0.4$. The root mean square error is 1.35 dB.

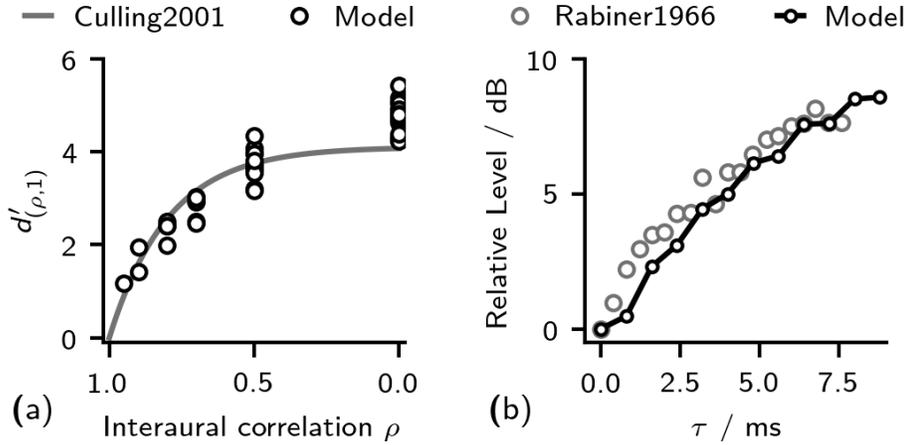

**Fig. 4: (a)** Comparison between the results of the correlation discrimination task of Culling et al. [36] (grey line shows their fit for the average subject) and the model performance in the same task **(b)** Comparison between the Model results (79 Hz filter bandwidth) and the data from Rabiner et al. [6] where the x-axis shows the group delay and the y-axis the associated threshold level relative to the $\tau = 0$ condition.

### C. Application of the model to literature data

The suggested model was additionally tested in comparison to literature data for interaural correlation discrimination [37] and binaural unmasking with arbitrary group delays [6, 39].

The current model can be directly applied to interaural correlation discrimination tasks. The internal noise was also kept unchanged. Results are shown in Fig. 4 (a), again with the model operating as an artificial observer. Following the approach of [37], the $d'$'s were calculated between selected interaural correlations as given in their Table 1. $d'$ for any of the measured values of interaural correlation $\rho$ and the fully correlated stimulus ($d'_{(\rho,1)}$) were then calculated by systematically summing over the calculated $d'$ as described in [37]. This process resulted in multiple approximations of $d'_{(\rho,1)}$ (e.g. $d'_{(0,0.8)} + d'_{(0.8,1)} = d'_{(0,1)}$ and $d'_{(0,0.5)} + d'_{(0.5,0.9)} + d'_{(0.9,1)} = d'_{(0,1)}$) all of which are given in Fig. 4 (a). It can be observed that the model is mostly able to reproduce the data. Only for uncorrelated noise (right-hand data points) the model slightly outperforms the average experimental data.

Additionally, the model was employed to simulate detection thresholds obtained with noise maskers with a time delay in the range of 0 to 7.8 ms and an additional phase shift to adjust the resulting interaural phase difference at the target frequency of 500 Hz to zero, i.e. a pure group delay [6]. Results are shown in Fig. 4b. While no adjustment of the model parameters was necessary to reproduce the data of [37], an increase in the standard deviation of the IPD noise to $\sigma_{\Delta\varphi} = 0.45$ rad was required here to simulate this data set. This increase in internal noise might be attributed to a shorter stimulus duration (while no quantification of the stimulus duration is given in [6] the stimuli were described as "short").

Thirdly, [39] measured tone detection thresholds in 50 Hz and 400 Hz wide noise maskers with a fixed group delay of 1.5 ms. Relative to their respective $N_0S_\pi$ reference, thresholds did not increase for the 50 Hz condition but they increased by about 5 dB in their 400 Hz condition. Their model could not account for their observation at all whereas the present model predicts a 2 dB increase for the 50 Hz condition and a 4 dB increase for the 400 Hz condition.

## IV. DISCUSSION

The experimental data and simulations presented in this study, assess the effect of tone detection thresholds as a combined function of noise delay and noise bandwidth. The lower the bandwidth of the noise and the higher thus the temporal coherence of the noise, the smaller the impact of time delay on the threshold. This relationship is confirmed by the data points at bandwidths smaller than the critical band. For larger stimulus' bandwidths the data does not depend on bandwidth. The experimental data systematically extends the classical delay dependence for broadband noise [2] by adding five narrower bandwidths of 25 to 200 Hz. It also extends [10] both in terms of bandwidths tested (6 instead of 2) and in terms of delay range (8 ms and infinity instead of 3 ms). The data set provides a valuable systematic baseline for testing binaural models.

In order to interpret the experimental results, a minimalistic phenomenological model based on the IPD model front-end [13] was employed. The most relevant aspects of the model are the monaurally-motivated 79-Hz wide 4[th]-order Gammatone filters [12] and the extraction of instantaneous IPDs. The model has two parameters: (1) a phase jitter, and (2) a detection uncertainty, both modelled by constant-variance internal noises. The model reproduces all trends in the data and can quantitatively account for all but the 25-Hz bandwidth data. Another minor deviation is that simulated thresholds still increase a little when increasing the bandwidth from 100 to 150 Hz, due to the energy in the filter tails. The experimental data flattens out marginally earlier. Both deviations would be reduced if a slightly narrower auditory filter was chosen. However, given the high accuracy obtained with the established 79-Hz ERB filter, we decided against adding an additional degree of freedom in the model evaluation.

The model uses the average absolute value of the instantaneous IPD as its decision metric. This metric is used with the underlying assumption that IPD fluctuations are a cue in binaural tone-in-noise detection tasks [34]. The binaural comparator element(s) can thus distinguish between, e.g. interaurally uncorrelated inputs causing large IPD fluctuations and a constant IPD of 90°. For a conventional cross-correlation element both stimuli generate the same output: zero correlation. For the present dataset, however, where all delayed noises have a long-term average IPD of zero, a simplified backend stage which only considers IPD deviations from zero is sufficient. This limitation makes the present version functionally equivalent to cross-correlation-coefficient based models and especially to the model of Bernstein and Trahiotis [10, 29], disregarding internal delays.

Bernstein and Trahiotis [29] accounted for almost all of the data from six seminal studies without requiring internal delays. They only employed internal delays for tone detection in delayed noise in contrast to the current study. In combination, [29] and the present study account for a total of nine data sets with an interaural coherence based model. It is beyond the scope of the present study to simulate data sets with non-zero mean IPDs, such as [2, 8], but first attempts with a manual setting of the mean IPD appear encouraging. Equivalently, correlation coefficient-based models could account for the binaural advantage offered by such noise, without requiring a delay line if they would use the complex-valued correlation coefficient, i.e. interaural coherence [11], rather than the real-valued correlation coefficient. Alternatively to the here suggested model approach, it can be assumed that the current data can also be explained using an imperfect delay line with p($\tau$) function and a similar data set could be simulated in such a model, e.g. [10]. Nevertheless, the simplistic IPD model of the present study indicates that a delay line is not required to account for the noise delay dependence of binaural unmasking.

The near-equivalence of the present IPD model and interaural correlation or coherence has further been shown exemplarily by simulating interaural correlation discrimination data [37] without changing model parameters. For the present model it does not make a difference if the masker-inherent IPD fluctuation, or synonymously the reduction in masker coherence, is introduced by a delay or by adding uncorrelated noise, which is in line with [2, 6]. Moreover, it was shown that the current model is able to account quantitatively for $S_\pi$ detection thresholds in noises with arbitrary group delays when the phase delay at 500 Hz is fixed at zero. For noises wider than the critical bandwidth it predicts a threshold increase of just under 3 dB per millisecond delay, a little more than observed in [6] and a little less than in 39]. For 50-Hz wide noise, the increase is closer to 2 dB/ms. Many delay-line-based approaches fail to account for this bandwidth dependence, because sensitivity is dictated by the bandwidth-independent delay line potency $\rho(\tau)$ [9, 39] rather than the bandwidth-dependent loss of coherence. However, the model employed in the present study is not expected to be the only model that can account for these data. All previous models that employ relatively narrow filters [10, 29] or a correspondingly steep $\rho(\tau)$ function [8] are expected to be isomorphic for the present data [40]. Our model is effectively the same as the analytic description and fitting from Rabiner et al. [6], their eq. 13 and, amongst others, conceptually identical to what was proposed by Langford and Jeffress [2]. Nevertheless, the present IPD-fluctuation-based model reflects a physiologically motivated concept to account for the delay dependence of tone detection with interaural phase differences using a minimum of model assumptions.

The model code has been made available in the auditory modeling toolbox.

## V.    CONCLUSIONS

The decline of binaural unmasking with increasing noise delay can be solely attributed to the noise coherence at the output of common auditory filters. This processing concept, first suggested by Langford and Jeffress [2], operates without any binaural-specific assumptions. An according minimalistic binaural model based on the physiologically plausible concept of extracting interaural phase fluctuations can be used to explain the data based on the monaurally derived auditory filter bandwidth without requiring delay lines. The results of the present study i) bridge between the mathematical concept of coherence and physiologically plausible auditory feature extraction and ii) help resolving the discrepancy between physiologic reports that cast doubt on neurons systematically compensating for larger delays [41-43] and psychoacoustically motivated models that appeared to require delay compensation [8-10, 29].

# Acknowledgements

This work was supported by the European Research Council (ERC) under the European Union's Horizon 2020 Research and Innovation Programme grant agreement no. 716800 (ERC Starting Grant to Mathias Dietz) and the Cluster of Excellence "Hearing4all" (DFG EXC2177/1) at the University of Oldenburg and the Deutsche Forschungsgemeinschaft DFG (SFB 1330 - 352015383 – Project A2 to Stephan Ewert and project B4 to Mathias Dietz)

# References

1. Hirsh: The influence of interaural phase on summation and inhibition. J. Acoust. Soc. Am. 20 (1948) 536-544.
2. T. L. Langford, L. A. Jeffress: Effect of noise crosscorrelation on binaural signal detection. J. Acoust. Soc. Am. 36 (1964) 1455–1458.
3. L. R. Bernstein, C. Trahiotis: Effects of interaural delay, center frequency, and no more than 'slight' hearing loss on precision of binaural processing: Empirical data and quantitative modeling. J. Acoust. Soc. Am. 144 (2018) 292–307. doi:10.1121/1.5046515
4. L. A. Jeffress: A place theory of sound localization. J. Comp. Physiol. Psychol. 41 (1948) 35–39.
5. C. E. Carr, M. Konishi: A circuit for detection of interaural time differences in the brain stem of the barn owl. J. Neurosci. 10 (1990) 3227-3246. doi:10.1523/JNEUROSCI.10-10-03227.1990

6. L. R. Rabiner, C. L. Laurence, N. I. Durlach: Further Results on Binaural Unmasking and the EC Model. J. Acoust. Soc. Am. 40 (1966) 62–70. doi:10.1121/1.1910065
7. R. M. Stern, G. D. Shear: Lateralization and detection of low-frequency binaural stimuli: effects of distribution of internal delay. J. Acoust. Soc. Am. 100 (1996) 2278–2288. doi:10.1121/1.417937.
8. M. van der Heijden, C. Trahiotis: Masking with interaurally delayed stimuli: the use of 'internal' delays in binaural detection. J. Acoust. Soc. Am. 105 (1999), 388–399.
9. J. Breebaart, S. Van De Par, A. Kohlrausch: Binaural processing model based on contralateral inhibition. I. Model Structure. J. Acoust. Soc. Am. 110 (2001) 1074–1088.
10. L. R. Bernstein, C. Trahiotis: Binaural detection as a joint function of masker bandwidth, masker interaural correlation, and interaural time delay: Empirical data and modeling," J. Acoust. Soc. Am. 148 (2020), 3481–3488. doi.org/10.1121/10.0002869
11. M. Dietz, G. Ashida: Computational Models of Binaural Processing, in Binaural Hearing. Litovsky RY, Goupell MJ, Fay RR, Popper AN, Editors New York, Springer. 2021 pp. 281-315. doi:10.1007/978-3-030-57100-9
12. B. R. Glasberg, B. C. J. Moore: Derivation of auditory filter shapes from notched-noise data. Hear. Res. 47 (1990) 103–138. doi:10.1016/0378-5955(90)90170-T
13. M. Dietz, S. D. Ewert, V. Hohmann: Auditory model based direction estimation of concurrent speakers from binaural signals. Speech Commun. 53 (2011) 592–605. doi:10.1016/j.specom.2010.05.006
14. M. Dietz, S. D. Ewert, V. Hohmann, B. Kollmeier: Coding of temporally fluctuating interaural timing disparities in a binaural processing model based on phase differences. Brain Res. 1220 (2008) 234–245. doi:10.1016/j.brainres.2007.09.026
15. P. X. Joris, B. van de Sande, A. Recio-Spinoso, M. van der Heijden: Auditory midbrain and nerve responses to sinusoidal variations in interaural correlation. J. Neurosci. 26 (2006) 279 –289.
16. R. Bamler, D. Just: Phase Statistics and Decorrelation in SAR Interferograms IGARSS '93. Better understanding of earth environment (1993) 980–984. doi: 10.1109/IGARSS.1993.322637
17. M. J. Goupell, W. M. Hartmann: Interaural fluctuations and the detection of interaural incoherence: Bandwidth effects. J. Acoust. Soc. Am. 119 (2006) 3971–3986. doi: 10.1121/1.2200147
18. M. M. Sondhi, N. Guttman: Width of the spectrum effective in the binaural release of masking. J. Acoust. Soc. Am. 40 (1966) 600–606.
19. A. J. Kolarik, J. F. Culling: Measurement of the binaural auditory filter using a detection task. J. Acoust. Soc. Am. 127 (2010) 3009–3017. doi:10.1121/1.3365314

20. S. van de Par, A. Kohlrausch: Dependence of binaural masking level differences on center frequency, masker bandwidth, and interaural parameters. J. Acoust. Soc. Am. 106 (1999) 1940–1947.
21. J. Breebaart, S. Van De Par, A. Kohlrausch: Binaural processing model based on contralateral inhibition. II. Dependence on spectral parameters. J. Acoust. Soc. Am. 110 (2001) 1089–1104.
22. T. Marquardt, D. McAlpine: Masking with interaurally "double-delayed" stimuli: The range of internal delays in the human brain. J. Acoust. Soc. Am. 126 (2009) EL177-EL182. doi:10.1121/1.3253689
23. M. Mc Laughlin, J. N. Chabwine, M. van der Heijden, P. X. Joris: Comparison of bandwidths in the inferior colliculus and the auditory nerve. II: Measurement using a temporally manipulated stimulus. J. Neurophysiol. 100 (2008) 2312–2327. doi:10.1152/jn.90252.2008
24. M. Mc Laughlin, B. van de Sande, M. van der Heijden, P. X. Joris: Comparison of bandwidths in the inferior colliculus and the auditory nerve. I. Measurement using a spectrally manipulated stimulus. J. Neurophysiol. 98 (2007) 2566–2579. doi:10.1152/jn.00595.2007
25. L. R. Bernstein, C. Trahiotis: Behavioral manifestations of audiometrically-defined 'slight' or 'hidden' hearing loss revealed by measures of binaural detection. J. Acoust. Soc. Am. 140 (2016), 3540–3548. doi: 10.1121/1.4966113
26. S. D. Ewert: AFC—A modular framework for running psychoacoustic experiments and computational perception models, in Proceedings of the International Conference on Acoustics AIA-DAGA2013, Merano, Italy. 2013 pp. 1326–1329.
27. H. Levitt: Transformed up-down methods in psychoacoustics. J. Acoust. Soc. Am. 49 (1971) 467–477.
28. S. Thavam, M. Dietz: Smallest perceivable interaural time differences. J. Acoust. Soc. Am. 145 (2019) 458–468.
29. L. R. Bernstein, C. Trahiotis: Converging measures of binaural detection yield estimates of precision of coding of interaural temporal disparities. J. Acoust. Soc. Am. 141 (2017) 1150–1160. doi:10.1121/1.4935606
30. W. T. Bourbon, L. A. Jeffress: Effect of bandwidth of masking noise on the detection of homophasic and antiphasic tonal signals. J. Acoust. Soc. Am. 37 (1965) 1180–1181.
31. V. Hohmann: Frequency analysis and synthesis using a Gammatone filterbank. Acta Acust. United Ac. 88 (2002) 433–442.
32. N. I. Durlach: Binaural signal detection: Equalization and cancellation theory, in Foundations of Modern Auditory Theory Vol. 2. Tobias J Editor New York, Academic Press. 1972 pp. 369–462.

33. S. D. Ewert, N. Paraouty, C. Lorenzi: A two-path model of auditory modulation detection using temporal fine structure and envelope cues. Eur. J. Neurosci. 51 (2020) 1265-1278. doi: 10.1111/ejn.13846
34. E. Zwicker, G. B. Henning: The four factors leading to binaural masking-level differences. Hear. Res. 19 (1985) 29–47. doi:10.1016/0378-5955(85)90096-6
35. M. J. Goupell, W. M. Hartmann: Interaural fluctuations and the detection of interaural incoherence. III. Narrowband experiments and binaural models. J. Acoust. Soc. Am. 122 (2007) 1029–1045. doi:10.1121/1.2734489
36. H. Lüddemann, H. Riedel, B. Kollmeier: Logarithmic scaling of interaural cross correlation: a model based on evidence from psychophysics and EEG, in Hearing: from sensory processing to perception – 14th International Symposium on Hearing. Kollmeier B, Klump G, Hohmann V, Langemann U, Mauermann M, Uppenkamp S, Verhey J Editors Berlin, Springer. 2007, pp. 379–388.
37. J. F. Culling, H.S. Colburn, M. Spurchise: Interaural correlation sensitivity. J. Acoust. Soc. Am. 110 (2001) 1020-1029. doi:10.1121/1.1383296
38. M. Dietz, J. H. Lestang, P. Majdak, R. M. Stern, T. Marquardt, S. D. Ewert, W. M. Hartmann, D. F. Goodman: A framework for testing and comparing binaural models. Hear. Res. 360 (2018) 92–106.
39. C. Trahiotis, L. R. Bernstein, M. A. Akeroyd: Manipulating the 'straightness' and 'curvature' of patterns of interaural cross correlation affects listeners' sensitivity to changes in interaural delay. J. Acoust. Soc. Am. 109 (2001) 321–330.
40. R. H. Domnitz, H. S. Colburn: Analysis of binaural detection models for dependence on interaural target parameters. J. Acoust. Soc. Am. 59 (1976) 598–601.
41. P. X. Joris, B. van de Sande, D. H. Louage, M. van der Heijden: Binaural and cochlear disparities. Proc. Nat. Acad. Sci. 103 (2006) 12917–12922.
42. D. McAlpine, D. Jiang, A. R. Palmer: A neural code for low-frequency sound localization in mammals. Nat. Neurosci. 4 (2001) 396–401.
43. S. K. Thompson, K. von Kriegstein, A. Deane-Pratt, T. Marquardt, R. Deichmann, T. D. Griffiths, D. McAlpine: Representation of interaural time delay in the human auditory midbrain. Nat. Neurosci. 9 (2006) 1096–1098.